\documentclass[%
 aip,
 amsmath,amssymb,
reprint,%
]{revtex4-1}

\usepackage[dvipdfmx]{graphicx}
\usepackage{epsfig}
\usepackage{url}

\usepackage{graphicx}
\usepackage{dcolumn}
\usepackage{bm}

\usepackage[utf8]{inputenc}
\usepackage[T1]{fontenc}
\usepackage{mathptmx}

\begin{document}

\preprint{AIP/123-QED}

\title{Application of single-electron effects to fingerprints of chips using image recognition algorithms}

\author{T. Tanamoto}
\affiliation{Corporate R \& D, Toshiba Corporation, Saiwai-ku, Kawasaki 212-8582, Japan} 
\altaffiliation[Present Address ]{Teikyo University.}
 \email{tanamoto@ics.teikyo-u.ac.jp}
\author{Y. Nishi}
\affiliation{Corporate R \& D, Toshiba Corporation, Saiwai-ku, Kawasaki 212-8582, Japan} 

\author{K. Ono}
\affiliation{Advanced Device Laboratory, RIKEN, Wako-shi, Saitama 351-0198, Japan}

\date{\today}

\begin{abstract}
Single-electron effects have been widely investigated 
as a typical physical phenomenon in nanoelectronics.
The single-electron effect caused by trap sites has been 
observed in many devices.
In general, traps are randomly distributed and not controllable;
therefore, different current--voltage
characteristics are observed through traps even in 
silicon transistors having the same device parameters (e.g., gate length).
This allows us to use single-electron effects as 
fingerprints of chips.
In this study, we analyze the single-electron effect of traps in conventional 
silicon transistors. 
At sufficiently low temperatures at which single-electron effects can be observed (in this case, 1.54 K),
we show that current--voltage characteristics can be used 
as fingerprints of chips through image recognition algorithms. 
Resonant tunneling parts in the Coulomb diagram can also be used supportively
to characterize each device in a low-temperature region.
These results show that single-electron effects 
can provide a quantum version of a physically unclonable function (quantum-PUF).
\end{abstract}

\maketitle

\begin{figure}
\centering
\includegraphics[width=4.5cm]{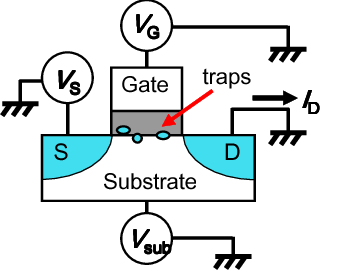}
\caption{
Schematic of metal-oxide-semiconductor field-effect transistor (MOSFET) channel with three trap
sites marked as small circles around the channel region. 
We measure the current between the source and the drain
as we vary the source-drain voltage ($V_S$), gate voltage ($V_G$), and substrate voltage ($V_{\rm sub}$). 
Trap sites can be treated as quantum dots having 
discrete energy levels.
}
\label{fig1}
\end{figure}

Single-electron tunneling (SET)~\cite{Averin,Kouwenhoven,Averin2,Mooij,Eiles,Sellier,Ono0,Lansbergen,Tan,Pierre,Gonzalez-Zalba,Schleser,Sigrist,Ono2} is experimentally observed when electrons are confined 
in a small space and the number of electrons is countable. 
Since the 1990s, many studies have used SET to investigate Kondo effects~\cite{Kastner,Park} and, more recently, quantum computing~\cite{Pla,Ono1,tanaQA1,tanaQA2}. 
Single-electron effects are also observed through trap sites in metal-oxide-semiconductor field-effect transistors (MOSFETs).
The single-electron effect caused by a trap site is understandable when trap sites are regarded as quantum dots (QDs), 
that is, a confined region of electrons with discrete energy levels (Fig.~\ref{fig1}). 
When a single QD is weakly coupled to both the source and the drain electrodes
and capacitively coupled to the gate electrode, the measurement of the drain current $I_D$ as a function of source voltage $V_S$ and gate voltage $V_G$ reveals a series of diamond-shaped regions where $I_D$ is strongly suppressed (see Fig.1 caption for measurement set up).

The features of the diamond-shaped regions, called Coulomb diamonds (CDs), indicate the threshold characteristics of SET.
The size of the CD measured in $V_S$ is the energy of Coulomb charging and/or quantum confinement for the dot. 
When more than two trap sites are coupled to the source and drain electrodes and contribute to SET, the CD shows a more complicated pattern comprising jagged corner lines~\cite{Ruzin}. 
Such complicated CD patterns represent the details of single-electron transport in multiple-QD devices~\cite{Kouwenhoven,Sellier,Ono1,Lansbergen}. 

Because the spatial positions, density, and energy levels of trap sites are not controllable, 
the features of CDs can differ in devices produced 
by the same fabrication process.
Therefore, one application of single-electron effects is the use of CDs as a ``fingerprint" of 
devices.

In general, the fingerprint of a device can be 
considered a physically unclonable function (PUF)
if it is unique, unclonable, and reliable. 
Emerging Internet of Things (IoT) technologies 
require a stable security system to protect users’ personal information.
PUFs are considered an important mechanism for 
providing a unique and inexpensive identification (ID) for each device.
A PUF outputs a response ID when a challenge signal is inputted
from a server or an authorized system.
PUF signals mostly originate from process variations of transistors
and circuits.
Basic static random access memory (SRAM) consists of 
two cross-coupled inverters and has fixed memory values 
only after a 0 or 1 datum is inputted. 
Therefore, the initial memory value is determined by the 
threshold variations of the transistors; this is the 
operating principle of SRAM-PUF~\cite{Guajardo,Holcomb}. 
The initial defects of memories~\cite{Marukame,RRAM} 
or the circuit delay~\cite{arbiter,Suh,wPUF} can also be used for 
generating PUFs.

Roberts {\it et al.} investigated PUFs using the quantum phenomena of 
resonant tunneling~\cite{Roberts}. 
\v{S}kori\'{c} proposed a quantum readout PUF 
in which a classical PUF is changed by the quantum state of photons
~\cite{Skoric}. 
Chen {\it et al.} proposed a PUF using traps in transistors (trap-PUF), 
in which the fingerprint of a chip is created by 
allocating 0 or 1 to each transistor depending on whether the trap site 
can be detected or not, respectively, in the range of the given voltage region~\cite{Chen}.
In the trap-PUF, when a 128-bit ID is requested, 
at least 128 transistors and related amplifying circuits are needed.
The individuality of chips in the trap-PUF is based  
on the fact that trap distributions cannot be controlled precisely
using current fabrication technologies.

In this study, we aim to further extend 
the use of trap sites in PUFs by using the 
single-electron effect of traps.
We use the features of CDs of traps as fingerprints of devices.
When we use single-electron effects as a PUF,
one transistor is expected to be sufficient 
for use as a fingerprint of the whole chip, 
thereby reducing the trap-PUF circuit area in the chip.
We further propose using image-matching algorithms~\cite{Prince} 
to identify different CDs of transistors. 
For this purpose, characteristic key points are abstracted by treating 
CDs as images as in conventional human
fingerprint detection. 
We demonstrate that this approach simplifies the identification process.

\begin{figure}
\centering
\includegraphics[width=8.6cm]{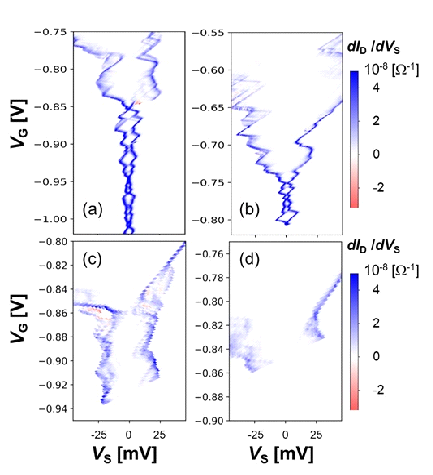}
\caption{
$dI_D/dV_S$ (differential conductance) characteristics of trap states in conventional 220-nm-wide transistors.
In our measurement, $|I_{D}| >$ 120 pA is not measured 
as values saturate in these regions. 
Thus, $dI_D/dV_S$ is nominally zero (white color) in these regions.
$dI_D/dV_S$ is 
calculated from measured current $I_D$ 
as a function of source voltage $V_S$.
The operating temperature in this study is 1.54 K.
(a) and (b) show the results for pMOSFETs having the same layout with 125-nm gate length.
(c) and (d) show the results for pMOSFETs having the same layout with 125-nm gate length and a 
silicon oxynitride gate dielectric.
Although these two devices have the same layout parameters ($L$ and $W$), 
their CDs show different characteristics and 
can therefore be used to identify each device.
} 
\label{fig2}
\end{figure}

\begin{figure}
\centering
\includegraphics[width=6.0cm]{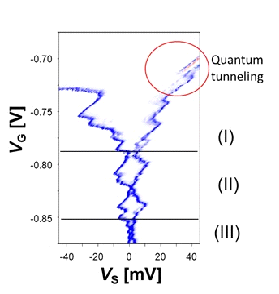}
\caption{
Most CD images can be separated into 
three regions based on the physical tunneling mechanism.
Region (I): Sub-threshold region. 
Region (II): Coulomb blockade caused by a couple of traps.
Region (III): Coulomb blockade caused by many traps. 
The blue part shows $dI_D/dV_S >0$. 
The red part shows $dI_D/dV_S <0$ where
we consider that resonant tunneling occurs.
The quantum region (resonant tunneling region) 
is mainly observed in region (I). 
}
\label{fig3}
\end{figure}

\begin{figure}
\centering
\includegraphics[width=8.6cm]{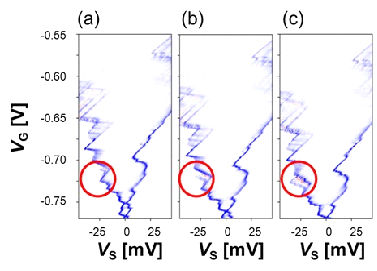}
\caption{
CD image changes when 
substrate voltage $V_{\rm sub}$ changes.
The differential conductance is measured 
in a conventional pMOSFET
with 220-nm width and 125-nm length.
(a) $V_{\rm sub}$ = 0.0 V, (b) $V_{\rm sub}$ = 0.05 V, 
and (c) $V_{\rm sub}$ = 0.1 V. 
The red circles show an example of a changed part
with different $V_{\rm sub}$.
}
\label{fig4}
\end{figure}

We prepared conventional p-type metal-oxide-semiconductor 
field-effect transistors (pMOSFETs) fabricated on our prototype production line.
These pMOSFETs have gate length $L$ = 125--145 nm 
and fixed gate width $W$ = 220 nm.
Coulomb blockade is observed at low temperature, and 
it disappears as the temperature increases to room temperature.
Figure~\ref{fig2} shows examples of the output of $dI_D/dV_S$ through trap sites
in pMOSFETs with $L$ = 125 nm.
As a reference, the average threshold voltage 
variations of transistors 
with $L$ = 125--135 nm are generally calculated in the range of 25--50 mV~\cite{Asenov}. 
Figure~\ref{fig2}(a) and (b) show the results for wafers with identical $L$ and $W$ that were simultaneously fabricated with the same process conditions. 
Figure~\ref{fig2}(c) and (d) show the results for wafers with identical $L$ and $W$ and an additional silicon oxynitride layer in the gate layer but fabricated
 with different process conditions. 
  
Trap sites, which are often defects, 
are generally removed to the greatest extent possible because 
they cause undesirable noises and nonuniformity in device characteristics.
These trap sites are constructed at an atomic level; 
therefore, their CDs are 
uncontrollable and unclonable by current CMOS technologies.
Thus, even if we fabricate transistors in the same product line, 
the features of the Coulomb blockade caused by single-electron effects 
can serve as a unique ID for each device.
A CD image is obtained by measuring the current $I_D$ as a function of gate voltage $V_G$ and 
source voltage $V_S$ and 
then numerically estimating the differential conductance $dI_D/dV_S$.

We divide the experimental data
into three regions based on their different physical mechanisms of Coulomb blockade.
As seen in Fig.~\ref{fig3}, CD images can be divided into three parts
depending on the magnitude of the gate bias.
Region (I): When $V_G$ is low, the current flows exponentially as a function of drain voltage. 
Region (II): When $V_G$ is applied beyond the subthreshold region, 
the current increases linearly as a function of $V_S$. 
Region (III): When $V_G$ is sufficiently large, the current saturates. 
Regions (I) and (III) correspond to the subthreshold and saturation region of 
conventional MOSFET operation, respectively.
In the subthreshold region, we can see the quantum aspect of the SET.
Without quantum effects, the $I_D$-$V_S$ characteristics show a step structure.
This results from classical Coulomb blockade in which the current is hindered 
until the applied voltage exceeds the potential at which the next energy 
level is occupied.

The CD images in Figs.~\ref{fig2} and \ref{fig3} have many similar jagged corner lines.
These features make identification 
difficult if we compare CD images numerically using conventional data 
such as $(V_S, V_G, dI_D/dV_S)$.
In ID applications, unstable output values are not desirable.
However, the output signals of devices do not always have identical values. 
Devices degrade owing to aging effects. 

The repetition of measurements between room temperature and low temperature 
might also change the device condition.
Thus, the current-voltage ($I$-$V$) characteristics 
will likely change every time the devices are measured.
The electronic states of QD devices are generally affected 
by both $V_{\rm sub}$ and $V_G$ values.
A change in $V_{\rm sub}$s changes the energy levels of trap sites 
and the carrier densities of electrodes, 
resulting in changes in CD images.
Therefore, we can emulate the effects of device 
conditions by changing $V_{\rm sub}$.

Figures~\ref{fig4} show CD images of a pMOSFET with $L$ = 135 nm and $W$ = 220 nm 
when the substrate bias $V_{\rm sub}$ changes from 0.0 to 0.1 V.
The details of the CD images clearly change with changes in $V_{\rm sub}$. 
Two transistors 
can be distinguished successfully if the two CD images can be 
distinguished under changes in $V_{\rm sub}$.
For example, in the detection of human fingerprints, the surface condition of human fingers changes depending on 
both internal and external conditions
such that human fingers are sometimes wet and sometimes oily.
However, fingerprints should be identified every time they are measured.
To identify fingerprints, key points in images of fingers are detected 
and compared with the image stored in the database.
Similarly, detecting 
key points should be effective for finding the similarity of CD images.
Although error-correction methods such as fuzzy algorithms can be used for this purpose~\cite{fuzzy}, 
we propose a direct method to find the similarity and difference 
by regarding measurement data as images. 
Then, experimental data can be compared more flexibly using images 
in a manner similar to comparing human fingerprints or pictures.

Many advanced recognition algorithms have been developed for feature detection and image matching over the years~\cite{Peng,Isik}.
Image matching software usually detect three image features: edge, corner, and flat. 
An edge is a line or border at which a surface terminates, 
a corner is a place where two converging lines or surfaces meet,
and a flat is a surface without any structures or marks.
We apply the AKAZE~\cite{AKAZE}, 
BRISK~\cite{BRISK}, and ORB~\cite{ORB}
recognition algorithms to obtain the key points of CD images.  
The AKAZE~\cite{AKAZE} algorithm uses a nonlinear diffusion filtering technique
whose scale spaces are constructed using a computationally efficient 
mathematical framework called fast explicit diffusion (FED).
ORB~\cite{ORB} is an extended algorithm using other algorithms 
that is rotation-invariant and noise-resistant. 
In the BRISK algorithm, key points are detected in octave layers of
the image pyramid as well as in in-between layers, 
and the sampling pattern consists of concentric circles 
in the neighborhood of each key point. 
In this study, we used 
Open Source Computer Vision Library (OpenCV ver.3)\cite{OpenCV}
based on Python 3.
The main advantage of using image matching software 
is that we can express the difference between two images 
by a single numerical value called $distance$ 
that is obtained as the output of each algorithm.

\begin{figure}
\centering
\includegraphics[width=8.2cm]{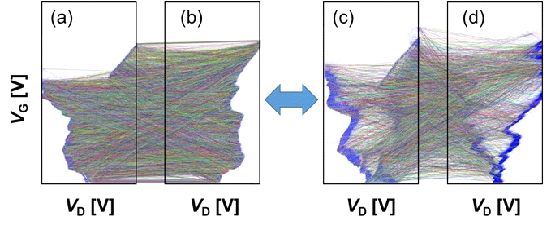}
\caption{
Examples of extraction of key points from CD images
in regions (I) and (II) using 
AKAZE recognition algorithm.
The distance between (a) and (b) is 46.52 ($V_{sub}$ = 0.5 V and 0.6 V)and 
that between (c) and (d) is 95.40 ($V_{sub}$ = 0.1 V and 0.14 V).
pMOSFETs with $L$ = 125 nm.}
\label{fig7}
\end{figure}
\begin{figure}
\centering
\includegraphics[width=7cm]{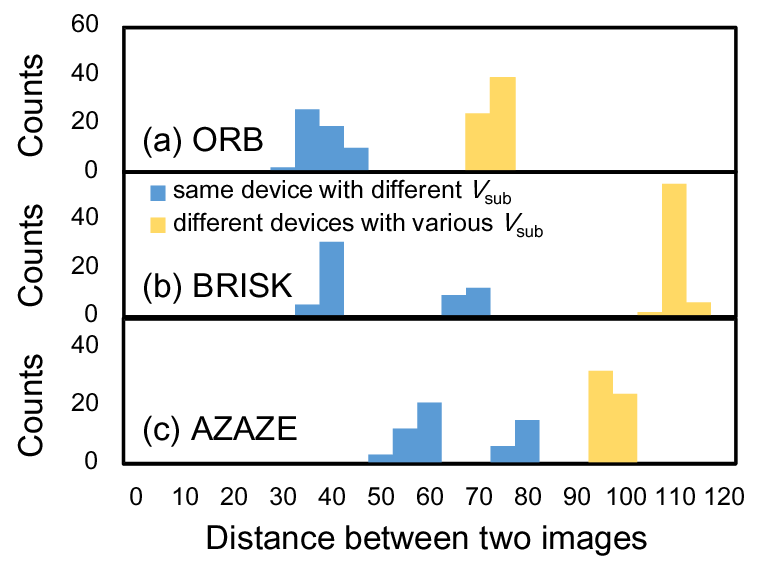}
\caption{
Comparison of histograms of distances of two images using three recognition algorithms
((a) ORB, (b) BRISK, and (c) AKAZE)
for 
changing $V_{\rm sub}$s. 
The scan region is restricted to 0.15 V. 
``Same device’’ shows 
the results for different $V_{\rm sub}$s with the same devices.
``Different device’’ shows the results for different $V_{\rm sub}$s 
with different devices.
}
\label{figC1}
\end{figure}
\begin{figure}
\centering
\includegraphics[width=7cm]{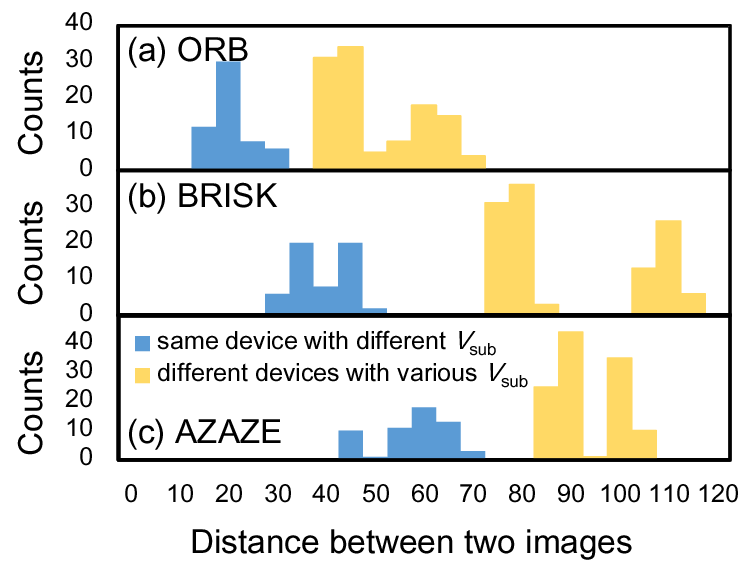}
\caption{Comparison of histograms of distances of two images using three recognition algorithms ((a) ORB, (b) BRISK, and (c) AKAZE) 
for 
changing $V_{\rm sub}$s. 
Devices with $L$ = 125 nm, $L$ = 135 nm, and $L$ = 140 nm are measured. 
}
\label{figC2}
\end{figure}

Figures~\ref{fig7} show an example of 
the extraction of characteristic key points between the two CD images of the same 
device ((a) and (b)) and different devices ((c) and (d)).
The connected lines between key points in Figs.~\ref{fig7}(a) and (b) 
look more condensed than those in Figs.~\ref{fig7}(c) and (d).
To understand the statistical characteristics,  
it is better to use histograms.
Figure~\ref{figC1} shows histograms of the $distances$ between the two CD images. 
The blue and yellow data show the distributions of the same and different devices, respectively, 
with different $V_{\rm sub}$s. 
The results show that the two devices can be distinguished by calculating the 
$distance$ of two CD images.
Each recognition algorithm has its own specific parameters 
such as a threshold and number of feature points.
As long as we chose several sets of parameters, 
we could not see any prominent improvement for some specific parameter sets.
Thus, we used the default parameters of each algorithm.
The two peaks in the distributions of the same devices 
(Figs.~\ref{figC1} (b) and (c)) are considered 
to originate from some detailed data structure in the same devices; 
we cannot explain the reason at present. 
Figure~\ref{figC2} shows histograms of the $distances$ between the two CD images for devices 
with different $L$s.
The $distances$ in devices with the same $L$s are clearly smaller 
than those in devices with different $L$s, 
and we can distinguish device IDs by using 
the CD images. 
At present, we could not judge which algorithm is best,
and therefore, it is better to use a couple of algorithms.

\begin{figure}
\centering
\includegraphics[width=8.6cm]{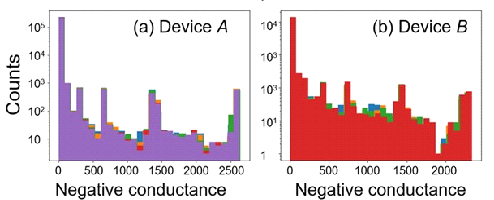}
\caption{
Histogram of differential conductance $dI_D/dV_S<0$ data in two devices 
$A$ and $B$ with 
$L$ = 125 nm and $W$ = 220 nm).
We can see differences between two devices even if they 
are made from the same transistors.
The different colors corresponds to various $V_{\rm sub}$s.
}
\label{fig9}
\end{figure}

Single-electron effects include various quantum tunneling processes
such as cotunneling~\cite{Averin2,Mooij,Eiles}.
Here, we simply investigate quantum effects where $dI_D/dV_S$ has negative values.
This is the result of resonant tunneling effects using the discrete
energy levels of trap sites. 
Because the resonant tunneling region is too small for using the 
image recognition algorithms, we use 
 histograms 
over data of negative differential conductance ($dI_D/dV_S<0$),
as shown in Fig.~\ref{fig9}. 
Each figure includes several distributions of the negative differential conductance
with different $V_{\rm sub}$s in the range of $\pm$0.1 V.
The histogram is divided into 50 regions in each of which 
values of the standard deviation are divided by their average.
The calculated deviation resulting from $V_{\rm sub}$ variations 
is 0.176\% for Fig.~\ref{fig9}(a) and 0.385\% for Fig.~\ref{fig9}(b). 
The relative difference of the two transistors is 
calculated by 
\begin{equation}
\frac{|{\rm Average(Fig.(a))}-{\rm Average(Fig.(b))}|}{
|{\rm Average(Fig.(a))}+{\rm Average(Fig.(b))}|/2}=0.83\%.
\end{equation}
Thus, the difference is not large. 
For increased effectiveness, methods such as that discussed in Ref.~\cite{Roberts} should be applied.
The present method can be applied easily and can be used supportively 
with the image recognition algorithms.
A detailed analysis of resonant tunneling will be conducted in future work.


In this study, CDs are measured at cryogenic temperatures. 
Note that single-electron effects can be observed 
even at room temperature if the device is designed appropriately~\cite{Uchida,Ahmed,Ray}.
However, this requires additional fabrication processes and incurs higher cost.
There is a trade-off between the fabrication cost and the operating temperature.
In this study, the number of tested transistors with the same $L$ and $W$
is limited because of the limited fabrication resources for wafers.
The main purpose of this study was to present the concept of 
a quantum PUF in single-electron devices. 
An examination of many transistors will be performed in a future study.

In summary, we have proposed a quantum PUF based on 
single-electron devices. 
In particular, we showed that 
we can distinguish two devices by using 
the distances calculated from CD images.
In contrast with the trap-PUF~\cite{Chen} 
in which many transistors need to be measured,
only one transistor is needed to generate the fingerprint of the chip,
and thus, the number of devices to be measured is reduced greatly. 
Note that electrons in traps can be treated as spin-qubits~\cite{Ono1}. 
Because trap distributions differ 
depending on the transistor, corresponding spin-qubit behaviors
are also expected to differ depending on the transistor.
Thus, the quantum behavior of an electron in a trap site 
can also be used as a fingerprint of a chip.
This will be explored in a future study.


\noindent
{\bf Acknowledgement}\\
TT thanks T. Hiraoka and T. Hioki for useful discussions.

\end{document}